\documentclass[pra,aps,twocolumn,amsmath,amssymb,superscriptaddress]{revtex4}
\usepackage{bm,graphicx,amsmath}
\usepackage{bbm}
\usepackage{bm}
\usepackage{amssymb}
\usepackage{subfigure}
\usepackage{epsfig}
\usepackage{epstopdf}
\usepackage{amstext}
\usepackage{latexsym}
\usepackage{times,palatino}
\usepackage{graphicx}

\usepackage{color}

\newcommand{\ket}[1]{\left\vert#1\right\rangle}
\newcommand{\bra}[1]{\left\langle#1\right\vert}

\newcommand{\qfi}{quantum Fisher information\,}
\newcommand{\rhor}{\tilde\rho}

\begin{document}

\title{Cavity-aided quantum parameter estimation in a bosonic double-well Josephson junction}

\author{ M. Zuppardo}\thanks{These authors contributed equally to this work}
\affiliation{Division of Physics and Applied Physics, Nanyang Technological University 637371, Singapore}
\author{J. P. Santos}\thanks{These authors contributed equally to this work}
\affiliation{Centro de Ci\^encias Naturais e Humanas, Universidade Federal do ABC, Santo Andr\'e, 09210-170 S\~ao Paulo, Brazil}
\affiliation{Centre for Theoretical Atomic, Molecular, and Optical Physics, School of Mathematics and Physics, Queen's University, Belfast BT7 1NN, United Kingdom}
\author{G. De Chiara}
\affiliation{Centre for Theoretical Atomic, Molecular, and Optical Physics, School of Mathematics and Physics, Queen's University, Belfast BT7 1NN, United Kingdom}
\author{M. Paternostro} 
\affiliation{Centre for Theoretical Atomic, Molecular, and Optical Physics, School of Mathematics and Physics, Queen's University, Belfast BT7 1NN, United Kingdom}
\author{F. L. Semi\~ao}
\affiliation{Centro de Ci\^encias Naturais e Humanas, Universidade Federal do ABC, Santo Andr\'e, 09210-170 S\~ao Paulo, Brazil}
\author{G. M. Palma}
\affiliation{NEST-INFM (CNR) and Dipartimento di Fisica e Chimica, Universit\`a degli Studi di Palermo, Via Archirafi 36, I-90123 Palermo, Italy}

\begin{abstract}
We describe an apparatus designed to make non-demolition measurements on a Bose-Einstein condensate (BEC) trapped in a double-well optical cavity. This apparatus contains, as well as the bosonic gas and the trap, an optical cavity. We show how the interaction between the light and the atoms, under appropriate conditions, can allow for a weakly disturbing yet highly precise measurement of the population imbalance between the two wells and its variance. We show that the setting is well suited for the implementation of quantum-limited estimation strategies for the inference of the key parameters defining the evolution of the atomic system and based on measurements performed on the cavity field. This would enable {\it de facto} Hamiltonian diagnosis via a highly controllable quantum probe. 
\end{abstract}

\date{\today}

\maketitle

The knowledge of the parameters entering the Hamiltonian of a given system is fundamental for a variety of tasks, from the formulation of accurate predictions on the behavior of the system to quantum state preparation and manipulation aimed at the achievement of an information processing goal. It is thus crucial to have the best possible characterisation of the key parameters entering the Hamiltonian of the system we would be interested, possibly in a weakly disturbing way for the dynamics that we aim at implementing. This is even more relevant for systems of difficult direct addressability or endowed with many mutually interacting degrees of freedom. This situation is very well embodied by intra-cavity atomic systems, whose potential for numerous applications of quantum information processing and quantum simulation has been affirmed by a series of ground-breaking experiments performed in recent years~\cite{ritsch,simulation}. In general, the determination of the features of a given model in these contexts requires measurements that are strongly disruptive for the fragile state of the system. A way around this problem is provided by the implementation of quantum non-demolition measurements~\cite{qnd}, which can be technically demanding.

In this respect, the approach based on the use of quantum probes of quantum evolutions, where a fully controllable probing device is coupled to the system of interest and subsequently measured to extract the relevant information, is very promising as it allows for the implementation of weakly disruptive strategies by means of indirect interrogation~\cite{probes}. Moreover, such an approach is prone to the application of sophisticated techniques for parameter estimation that aim at determining the best preparation and measurement of the probe and are explicitly designed to achieve the best possible accuracy of estimation allowed by classical and quantum mechanics~\cite{parisReview}. 

In this paper, we move along these interesting lines and propose the use of quantum estimation techniques (QET) to determine the crucial parameters entering a system consisting of a cold atomic ensemble loaded into a double-well potential. Our probing system is embodied by the field of an optical cavity that is locally coupled to only one well of the potential, along the lines of earlier proposals~\cite{homodyne,w-milbourn}. We show that strategies for both the sequential and simultaneous estimation of both the tunnelling rate and the on-site repulsion energy can be successfully applied to gather key information on the evolution that the atomic system would undergo. Moreover, using matter-to-light mapping techniques, we find the conditions that allow for the non-disruptive determination of the population imbalance between the wells by means of measurements performed only on the cavity field.

The remainder of this paper is organised as follows. In Sec.~\ref{model} we introduce the system that we aim at studying and perform a basic analysis of its dynamical features. In Sec.~\ref{mapping} we describe a simple matter-to-light map that allows us to infer the population unbalance between the two wells under conditions of weak cavity-atom coupling. Sec.~\ref{QET} is devoted to the description of the quantum estimation theory method that we implement in order to infer the parameters of the Hamiltonian model regulating the dynamics of the atom-loaded double well. Finally, in Sec.~\ref{conc} we draw our conclusions and highlight a few open directions of investigation.

\section{The model}
\label{model}

As shown in Fig.~\ref{sistema}, our system consists of a Bose-Einstein condensate of two-level atoms trapped in a double-well potential. As introduced above, one of the wells is accommodated within a single-mode  optical cavity pumped by an external laser field. The frequency of the field $\omega_C$ is assumed to be much different than the atomic transition frequency, so that the excited state of the atoms can be adiabatically eliminated from the dynamical picture. In such a dispersive-interaction regime, the condensate acts as a quantum dielectric medium for the cavity field, modifying its refractive index.

\begin{figure}[!t]
\label{sistema}
\centering
  \includegraphics[width=7 cm]{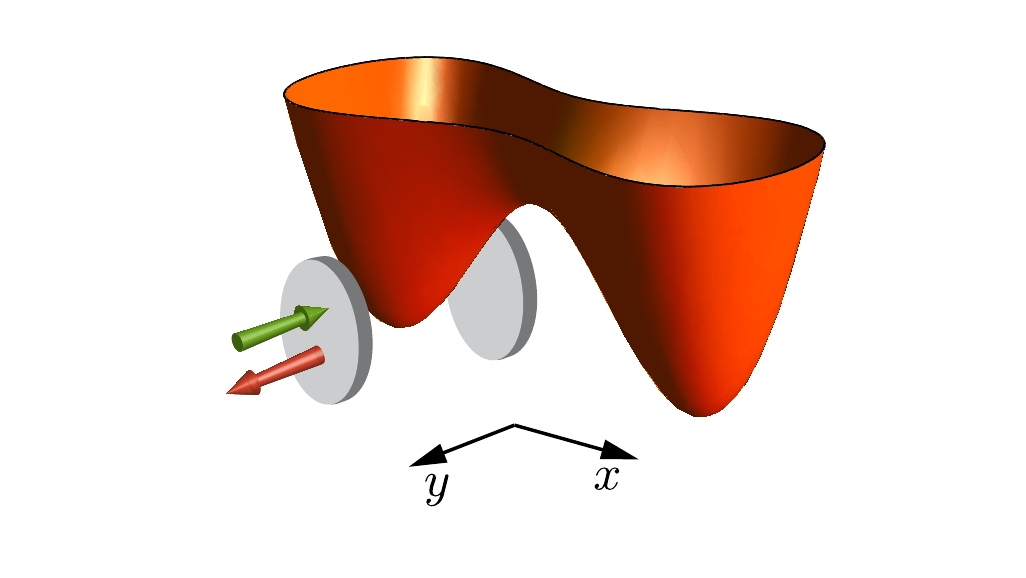}
  \caption{(Color online) Schematic representation of the system. A double-well potential loaded with ultra cold atoms is coupled to an optical cavity that accommodated only one of the wells. The cavity is pumped by a classical field and the signal leaking out of the resonator is measured to infer the key features of the double-well system.}\label{sistema}
\end{figure}
The total system Hamiltonian is thus $\hat H=\hat H_{A}+\hat H_{C}+\hat H_{I}$, where
$ \hat H_C=\omega_C \hat a^\dagger \hat a$ is the energy of the cavity field [we use units such that $\hbar=1$ throughout this paper], $\hat H_A$ is the energy of the atomic system, and $\hat H_I$ is the matter-light interaction term. Using the so-called two-mode approximation (TMA) introduced in Ref.~\cite{w-milbourn}, the atomic energy can be written as 
\begin{equation}
\label{hat}
\hat H_A=E_0(\hat c_1^\dagger \hat c^{{}}_1+ \hat c_2^\dagger \hat c^{}_2)+ \kappa(\hat c_1^{\dagger 2}\hat c_1^2+\hat c_2^{\dagger 2}\hat c_2^2)+R(\hat c_1^\dagger \hat c_2+\hat c_2^\dagger \hat c_1).
\end{equation}
Here $\hat c_{1,2}$ are the bosonic operators that annihilate an atom in its ground state in each of the two wells used to describe the state of the atomic system in the double well within the TMA, $E_0$ is the energy of the single atom in one of the wells, $\kappa$ represents the strength of the interaction among the atoms (about $4\times 10^{-34}$ J for a gas of $^{87}$Rb), and $R$ is a tunnelling rate proportional to the probability that an atom goes from a well to the other and dependent on the height of the barrier. 
All these terms are responsible for the free evolution of the system, which we are interested in monitoring. As discussed in Ref.~\cite{oberthaler}, based on the ratio between $R$ and $\kappa$, several regimes can be distinguished in the evolution of the atoms. For example, when $R\ll \kappa$, the atoms undergo a regime known as self-trapping, in which the tunnelling is almost completely suppressed and the difference in population between the wells stays almost constant in time.

As for the cavity-atom interaction terms, in the regime relevant to our work this takes the form~\cite{brennecke1}
\begin{equation}
\hat H_{I}=\beta \hat c_1^\dagger \hat c_1\hat a^\dagger \hat a,
\end{equation}
where $\beta=\frac{1}{\Delta_a}\int G (\vec r) |u_1(\vec r)|^2 d^3 r$, with $\Delta_a$ representing the detuning between the frequency of the atomic transition and that of the cavity,  $u_1(\vec r)$ the eigenfunction of the single-particle Hamiltonian of the first well, and $ G (\vec r)$ is a factor proportional to the intensity of the field inside the cavity.
We note that the interaction term is inversely proportional to the detuning $\Delta_a$. This value can be tuned by modifying the frequency of the cavity, varying its geometrical size or by working on the atomic spectrum, making a convenient use of the Zeeman or Stark effect. The interaction can thus be tuned over quite a large range of values. 
Let us consider how the state of the cavity, which evolves under the effect of the Hamiltonian $\hat H_C+\hat H_I$, is affected by the atomic dynamics.

By taking in consideration the leaky nature of any realistic cavity, the overall dynamics of the system density matrix $\rho$ is well described by a master equation reading, in a frame rotating at the cavity-field frequency $\omega_C$, as
\begin{equation}
\label{lindblad3}
\begin{aligned}
\frac{d}{dt} \rho =&-i\beta[\hat a^\dagger \hat a\hat n_1, \rho]-i[\hat H_A, \rho]-\gamma{\cal L}(\rho)
\end{aligned}
\end{equation}
with ${\cal L}(\rho)=(1+N_c)D[\hat a,\rho]+ N_cD[\hat a^\dag,\rho]$ 
and  $D[\hat O,\rho]=([\hat O^\dagger \hat O, \hat \rho]_+/2-\hat O\hat \rho \hat O^\dagger)$ for any operator $\hat O$. Here, $\gamma$ is the decay rate of the cavity, $\hat n_1=\hat c_1^\dag\hat c_1$, and $[\cdot,\cdot]_+$ the anticommutator. At optical frequencies, the number of thermal environmental photons $N_c$ is very small. We can thus take $N_c\simeq 0$ and solve Eq.~\eqref{lindblad3} numerically with the quantum jump method~\cite{Dalibar}. 

\begin{figure}[t]
\centering%
  \includegraphics[width=\columnwidth]{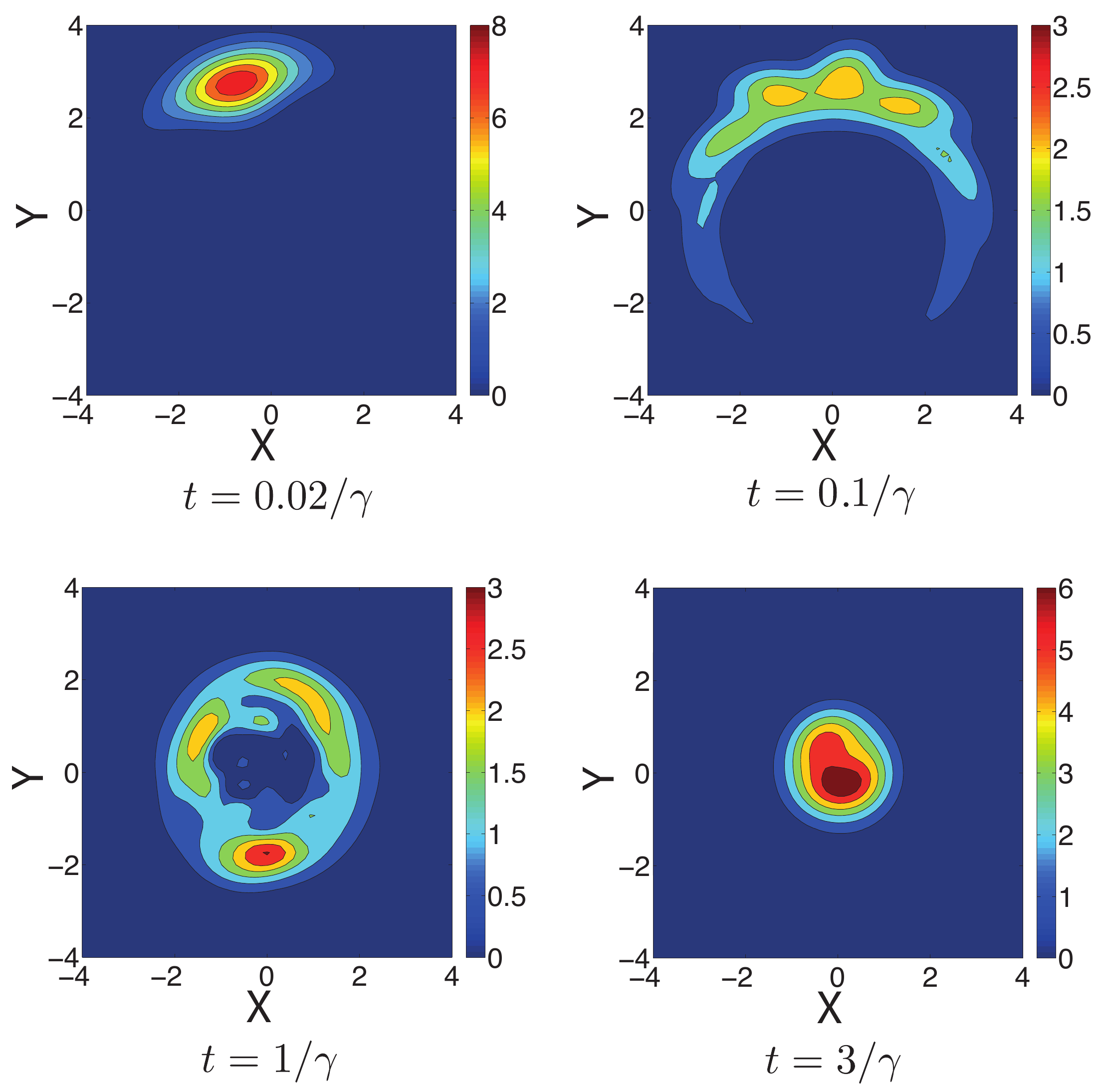}
\caption{(Color online) Evolution of the Wigner function of a high-Q cavity, interacting with a BEC in a double-well, out of the self-trapping regime. The parameters value are $\beta=\kappa=R\simeq 8\times 10^{-37}\gamma$ and $N=30$.
The initial state is coherent  with amplitude $\alpha=1.5$.}
\label{wigner}
\end{figure}

An informative indicator of the resulting dynamics comes from the study of the temporal behavior of the Wigner function of the cavity field
\begin{equation}
W(x,p)=\frac{1}{\pi^2}\int\text{Tr}[\rho_C(t)e^{y\hat a^\dag-y^*\hat a}]e^{2i(y_rp-y_ix)}d^2y
\end{equation}
with $\rho_C(t)=\text{Tr}_A[\rho(t)]$ the reduced state of the cavity field after tracing out the condensate's degrees of freedom, $(x,p)$ the phase-space variables, and $y=y_r+iy_i$.
The results of our simulations are shown in Fig.~\ref{wigner}, where we have considered the cavity field as initially prepared in a coherent state $\ket{\alpha}~(\alpha\in{\mathbb R}$ for simplicity) and the condensate in the Fock state $\ket{n_1,n_2}$. We see that $W(x,p)$ evolves into a ring with a radius  of about $|\alpha|$. From different calculations, we found that this effect is also visible if the gas is in the self-trapping regime. However, in this case, the ring shape is reached in  much longer times.
We can give a qualitative interpretation of this behavior. The quantity $\beta \hat n_1$ is added to $\omega_c$, acting like an effective frequency for the cavity. 
This frequency, however, is not a number but an operator acting on the Hilbert space of the gas, which has a discrete spectrum that is upper-bounded by the total number of atoms $N$. Each of these eigenvalues has a certain probability $|\langle n_1|\hat\rho|n_1\rangle|^2$ to occur, which depends on the state of the atoms at that time. 
Each eigenvalue is effectively a frequency of the cavity, and so it has the effect of making the initial Wigner function rotate with angular speed $\beta n_1+\omega_c$. Hence we can say that  the Wigner function for $t>0$ is the superposition of many Gaussians, which rotate at different speeds. At very large times the cavity decays and the Wigner function is not distinguishable from the one of the vacuum state.
In a frame rotating at speed $\omega_c$, the interaction with the atoms is the only term responsible for the evolution, apart from the external noise. In the high-Q limit, then, the cavity field is indeed very sensitive to the interaction with the atoms, even at small $\beta$.

On the other hand, the interaction term obviously also affects the dynamics of the atoms. However, it is natural to think that all the effects of the interaction on the atoms are less relevant if $\beta$ is sufficiently small compared to $\kappa$ and $R$. As our purpose is to measure observables of the condensate in a weakly invasive way, we want to operate in regimes in which $\beta\ll R,k$. We will use this assumption in the next Section.

\section{Matter-to-light mapping to infer the population imbalance}
\label{mapping}

 In this regime, measurements of the transmitted or scattered light can then be used to implement a quantum non-demolition measurement of the state of the condensate~\cite{mekhov}. In this spirit, here we will illustrate a scheme in which the dispersive coupling of an atomic condensate trapped in a double well potential is used to perform a weak measurement of the number of atoms --- and of their fluctuations --- trapped in the potential. 
 
\begin{figure*}[!ht]
  {\bf (a)}\hskip7cm{\bf (b)}\\
\includegraphics[width=\columnwidth]{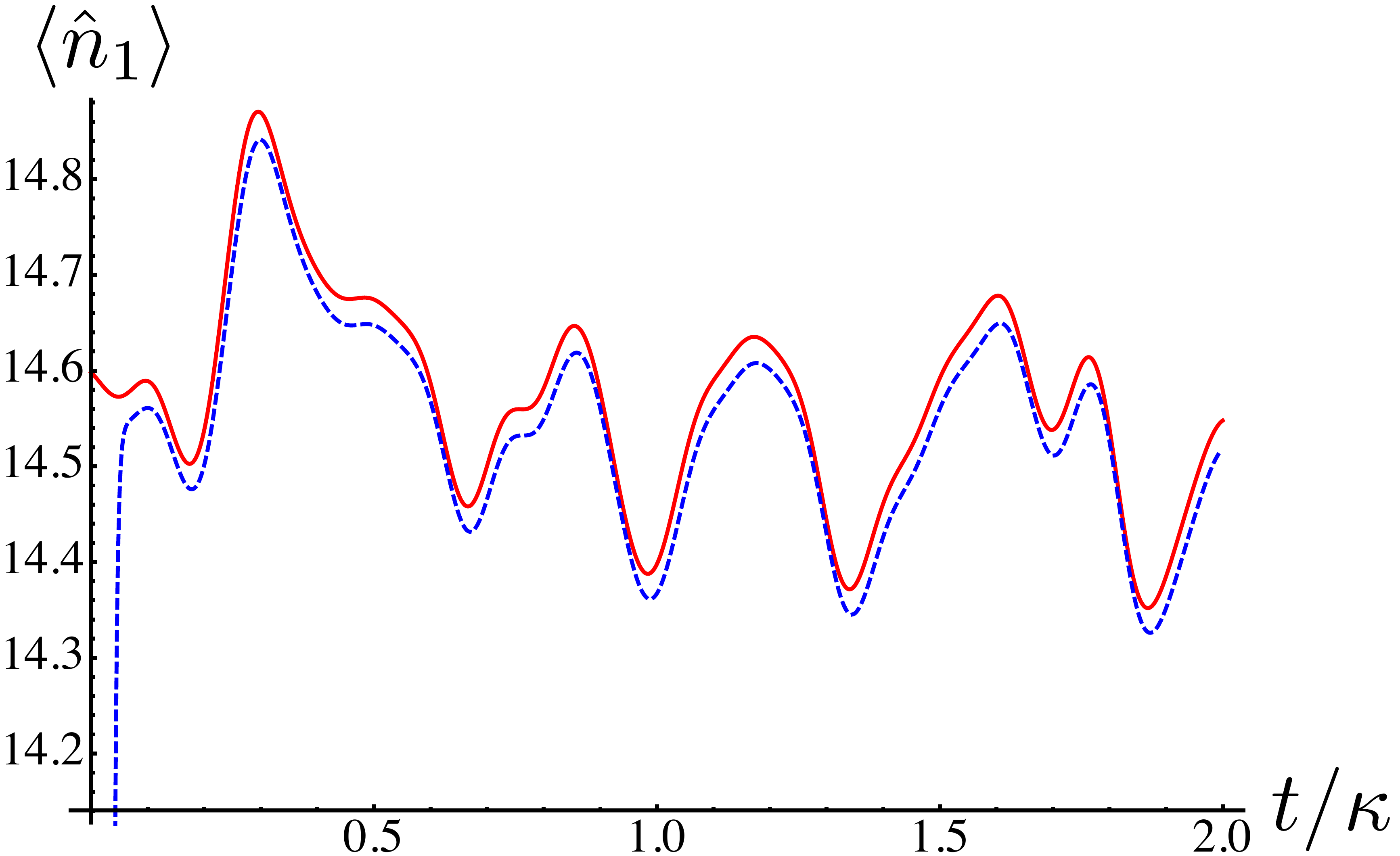}\includegraphics[width=0.95\columnwidth]{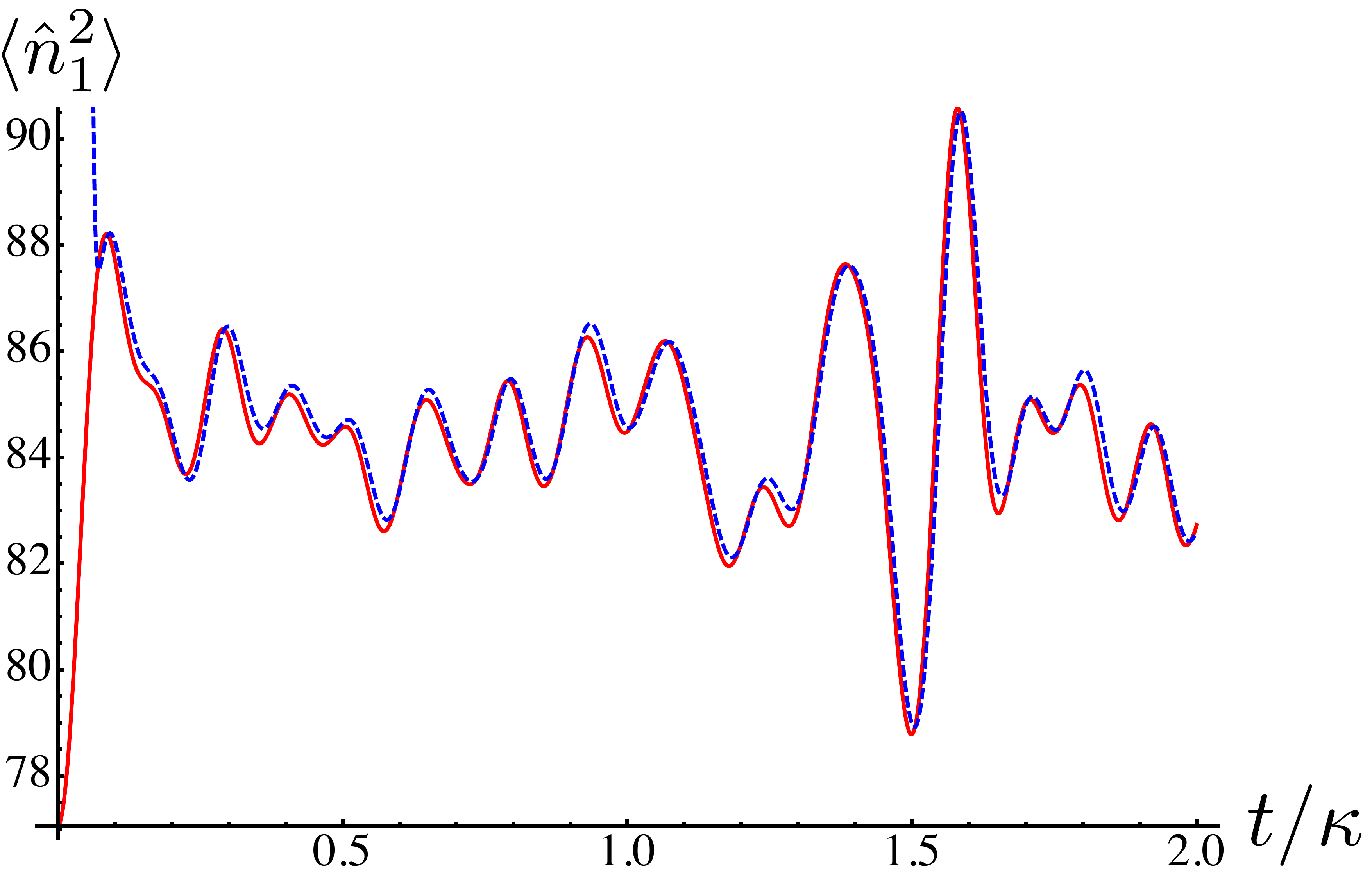}\\
  {\bf (c)}\hskip7cm{\bf (d)}\\
\includegraphics[width=\columnwidth]{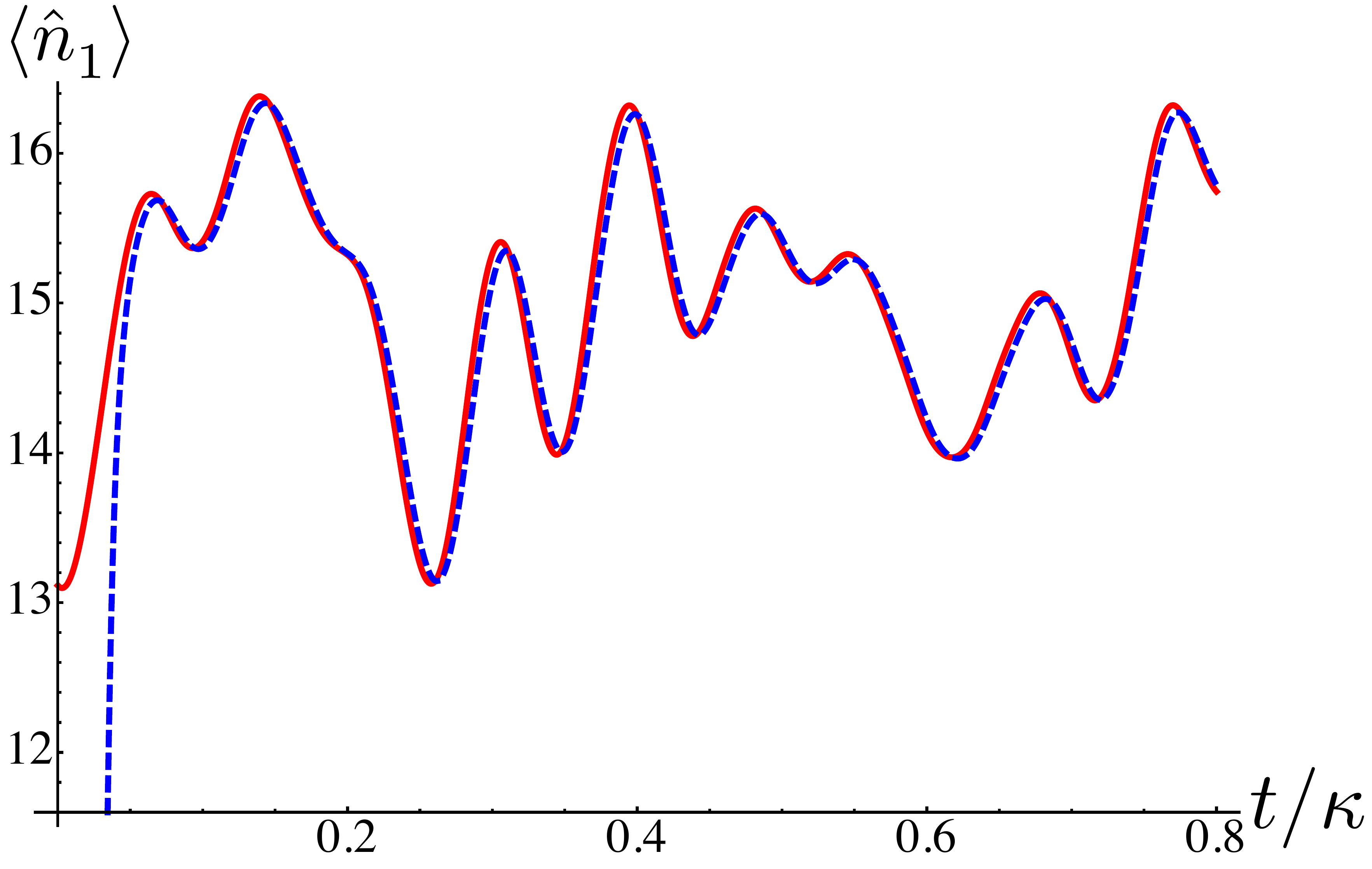}\includegraphics[width=0.95\columnwidth]{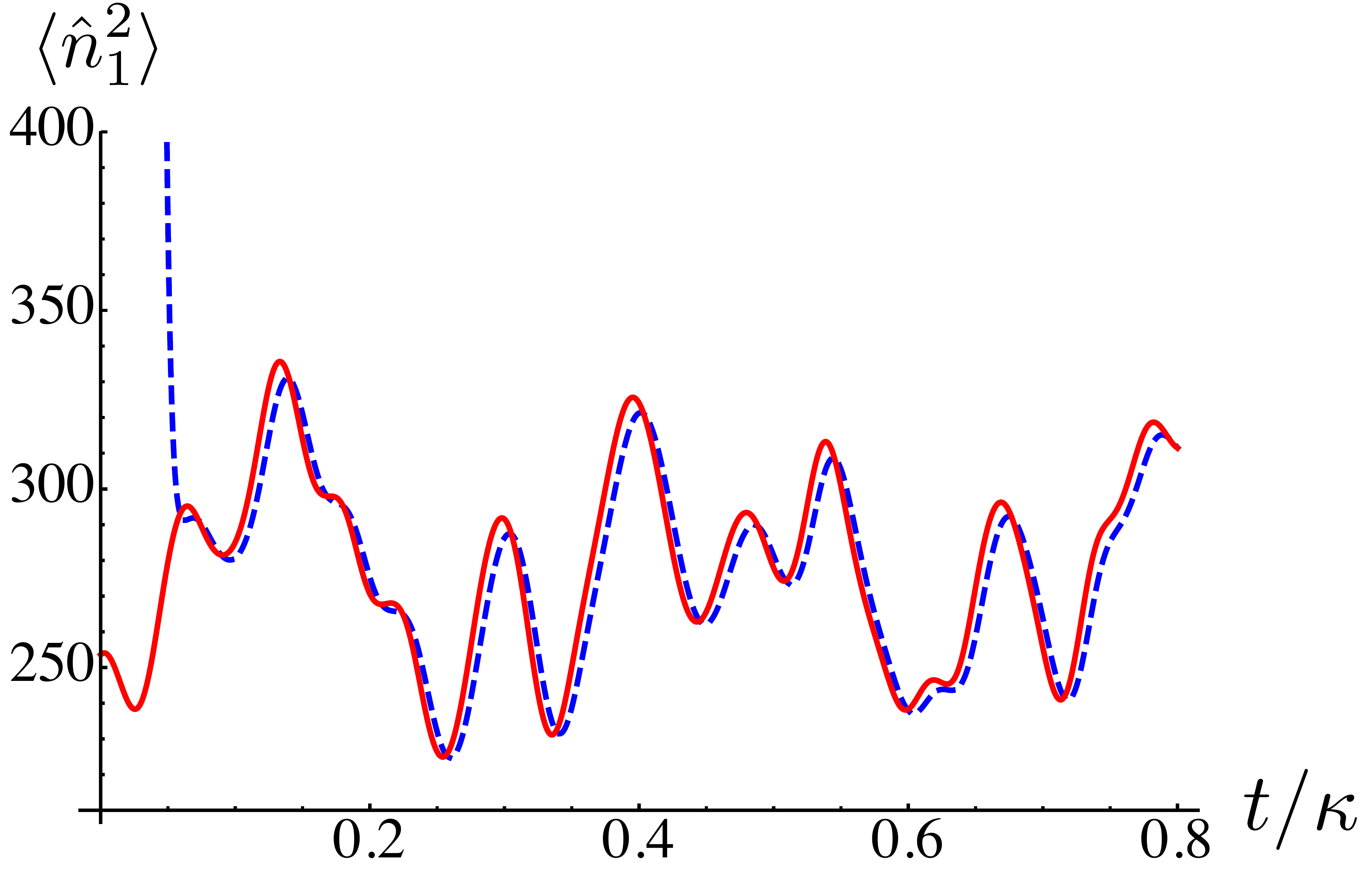}
  \caption{(Color online) Time-dependence of $\langle \hat n_1\rangle$ [panel {\bf (a)} and {\bf (c)}] and $\langle\hat n_1^2\rangle$ [panel {\bf (b)} and {\bf (d)}]. The solid red line represents their numerical evolution, while the blue dashed lines show the results of numerical estimations obtained using Eqs.~\eqref{n1m} and \eqref{varn1}, and the parameters $N=30, \eta=\kappa\simeq 4\times 10^{-34}\text{J}$, $\gamma/\kappa=500$, $\beta/\kappa=1/16$, which are valid for a system of $^{87}$Rb atoms. In panels {\bf (a)} and {\bf (b)} we have used a value of the tunnelling rate comparable to the self-interaction energy ($R/\kappa=1$), while panels {\bf (c)} and {\bf (d)} are for a tunnelling dominated situation ($R/\kappa=30$).} 
\label{nm}
\end{figure*} 
 
In order to establish the formal link between the cavity-field operators to the atomic ones, we consider the cavity as pumped by an external coherent field at frequency $\omega_p$, so that an extra term of the form $\hat H_{P}=\eta(\hat a^\dag e^{-i\omega_p t}+h.c.)$ (with $\eta$ the cavity-pump interaction strength) is added to the Hamiltonian model $\hat H$. Moving to a frame rotating at the frequency of the pump, the dynamics of the cavity field is effectively captured by considering the Langevin equation 
\begin{equation}
\label{aina}
\partial_t \hat a=-i(\Delta_C +\beta\hat n_1)\hat a+\sqrt{\gamma}\hat a_{in}(t)-\frac{\gamma}{2}\hat a(t)+\eta\hat\openone,
\end{equation}
where $\Delta_C=\omega_C-\omega_p$ is the detuning between the pump and the cavity, $\gamma$ is the single-photon damping rate of the cavity, $\hat a_{in}(t)$ is the annihilation operator describing input noise to the cavity.
We assume $\gamma$ to be so large that the cavity field reaches a steady-state in a time that is much shorter than the evolution of the atomic system and the typical timescale $\beta^{-1}$ of the cavity-atom interaction. If so, the system will be in its steady state after a time of the order of $\gamma^{-1}$. By setting $\partial_t \hat a=0$ in Eq.~\eqref{aina} and solving with respect to $\hat a$, we get
\begin{equation}
\hat a=\frac{\eta\hat\openone+\sqrt{\gamma}\hat a_{in}(t)}{i\beta\hat n_1+{\gamma}/{2}}.
\end{equation}
We also assume that the pump is resonant with the cavity, $\Delta_C=0$ (this assumption is not essential, but it greatly simplifies the results).
We already assumed $\beta$ to be small and $\gamma$ to be large.  We now also request $\beta N\ll\gamma$ and we expand the ratio up to the second order in $2\beta \hat n_1/\gamma$ to get
\begin{equation}
\hat a=\frac{2\eta+2\sqrt{\gamma}\hat a_{in}(t)}{\gamma}\left(\hat\openone-\frac{2i\beta \hat n_1}{\gamma}-\frac{4\beta^2\hat n_1^2}{\gamma^2}\right).
\end{equation}
The cavity quadrature operators are then
\begin{equation}
\label{pp}
\hat P=-\frac{4\sqrt{2}\eta\beta}{\gamma^2}\hat n_1+\frac{2}{\sqrt{\gamma}}\left[\hat P_{in}(t)\left(\openone-\frac{4\beta^2\hat n_1^2}{\gamma^2}\right)-\hat X_{in}\frac{2\beta\hat n_1}{\gamma}\right],
\end{equation}
and
\begin{equation}\label{xx}
\begin{split}
\hat X &  = \frac{2\sqrt{2}\eta}{\gamma}\left(1-\frac{4\beta^2}{\gamma^2}\hat n_1\right)\\&+\frac{2}{\sqrt{\gamma}}\left[\hat X_{in}(t)\left(\openone-\frac{4\beta^2\hat n_1^2}{\gamma^2}\right)+\hat P_{in}\frac{2\beta\hat n_1}{\gamma}\right].
\end{split}
\end{equation}
These quantities can be measured by homodyne measurements of the output field. If we assume white noise entering the cavity (i.e. a zero-mean, Delta-correlated field), from the last relations we obtain 
\begin{equation}
\label{p}
\langle\hat P\rangle=-\frac{2\sqrt{2}\eta}{\gamma}\left(\frac{2\beta \langle\hat n_1\rangle}{\gamma}\right),~~~~\langle\hat X\rangle=\frac{2\sqrt{2}\eta}{\gamma}\left(1-\frac{4\beta^2\langle\hat n_1^2\rangle}{\gamma^2}\right).
\end{equation}
In order to get these results we have considered no correlations between the input field and the atoms. By inverting Eq.~\eqref{p}, we get a relation for the number of atoms in the first well
\begin{equation}
\label{n1m}
\langle\hat n_1\rangle=-\frac{\gamma^2}{4\sqrt{2}\beta\eta}\langle \hat P\rangle.
\end{equation}
Using the second of Eqs.~\eqref{p} we obtain
\begin{equation}
\label{varn1}
\langle\hat n_1^2\rangle=\frac{\gamma^3}{8\sqrt{2}\beta^2\eta}\left(\frac{2\sqrt{2}\eta}{\gamma}-\langle \hat X\rangle\right).
\end{equation}
The last two equations give a way to determine the average number of atoms and its variance, through measurements of the mean value of the quadrature operators of the output field.
In order to make these two equations describe the dynamics of the condensate, we require
\begin{equation}
 \gamma\gg(\beta N, \kappa N, R)
 \end{equation}
Furthermore, we want that such measurement does not perturb the atoms strongly. If we add the assumption
\begin{equation}
\beta \langle \hat a^\dagger\hat a(0)\rangle\ll\kappa N, R,
\end{equation}
we expect the free dynamics of the atoms to be predominant. To verify the validity of such results, we solve Eq.~\eqref{lindblad3} numerically to track the system's dynamics and build a suitable benchmark. Figs.~\ref{nm} shows the results of such simulations. We see that, after a transient, the field follows closely the evolution of the atoms, whose dynamics is not significantly different from the free one, which certifies the weakly disturbing nature of the probing mechanism at hand. Quite evidently, the validity of such adiabatic following holds true regardless of the conditions of the atomic evolution. In fact, while panel {\bf (a)} and {\bf (b)} address the case of a tunnelling rate comparable with the atomic self-interaction energy, panels {\bf (c)} and {\bf (d)} displays the results valid for a tunnelling-dominated regime ($R=30\kappa$). A feature that is common to all of the simulations that we have produced is the small time-delay between the actual dynamics of ($\langle\hat n_1\rangle, \langle\hat n^2_1\rangle$) and their estimations achieved through the light-matter mapping. This is not a numerical artefact but actually describes the fact that the cavity needs to reach a dynamical steady state in order to adapt to the dynamics of the atoms. In fact, the time delay of the red curves with respect to the blue ones in Fig.~\ref{nm} is of the order of $\gamma^{-1}$.

\begin{figure}[b]
{\bf (a)}\hskip3.5cm{\bf (b)}
  \includegraphics[width=0.9\columnwidth]{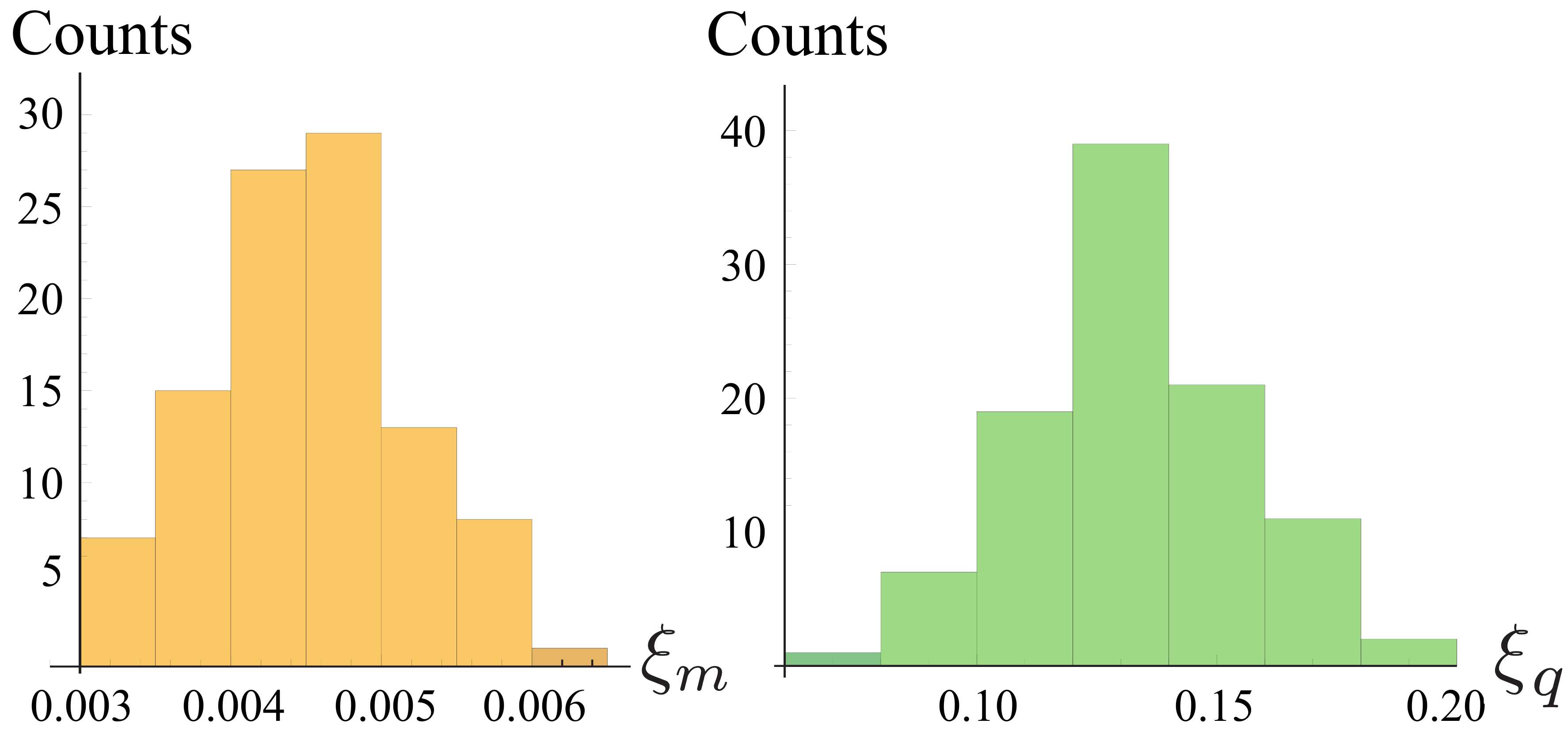} 
    \caption{(Color online) Distribution of values taken by the parameter $\xi_m$ [panel {\bf (a)}] and  $\xi_v$ [panel {\bf (b)}] obtained for 100 randomly-generated initial states of the BEC. Here we have taken the parameters used in panels {\bf (a)} and {\bf (b)} of Fig.~\ref{nm}. We have taken $t_0=0.07/\kappa$ and $t_1=0.8/\kappa$.}
  \label{ximxiv}
\end{figure}

\begin{figure}[b]
{\bf (a)}\hskip3.5cm{\bf (b)}
\includegraphics[width=0.55\columnwidth]{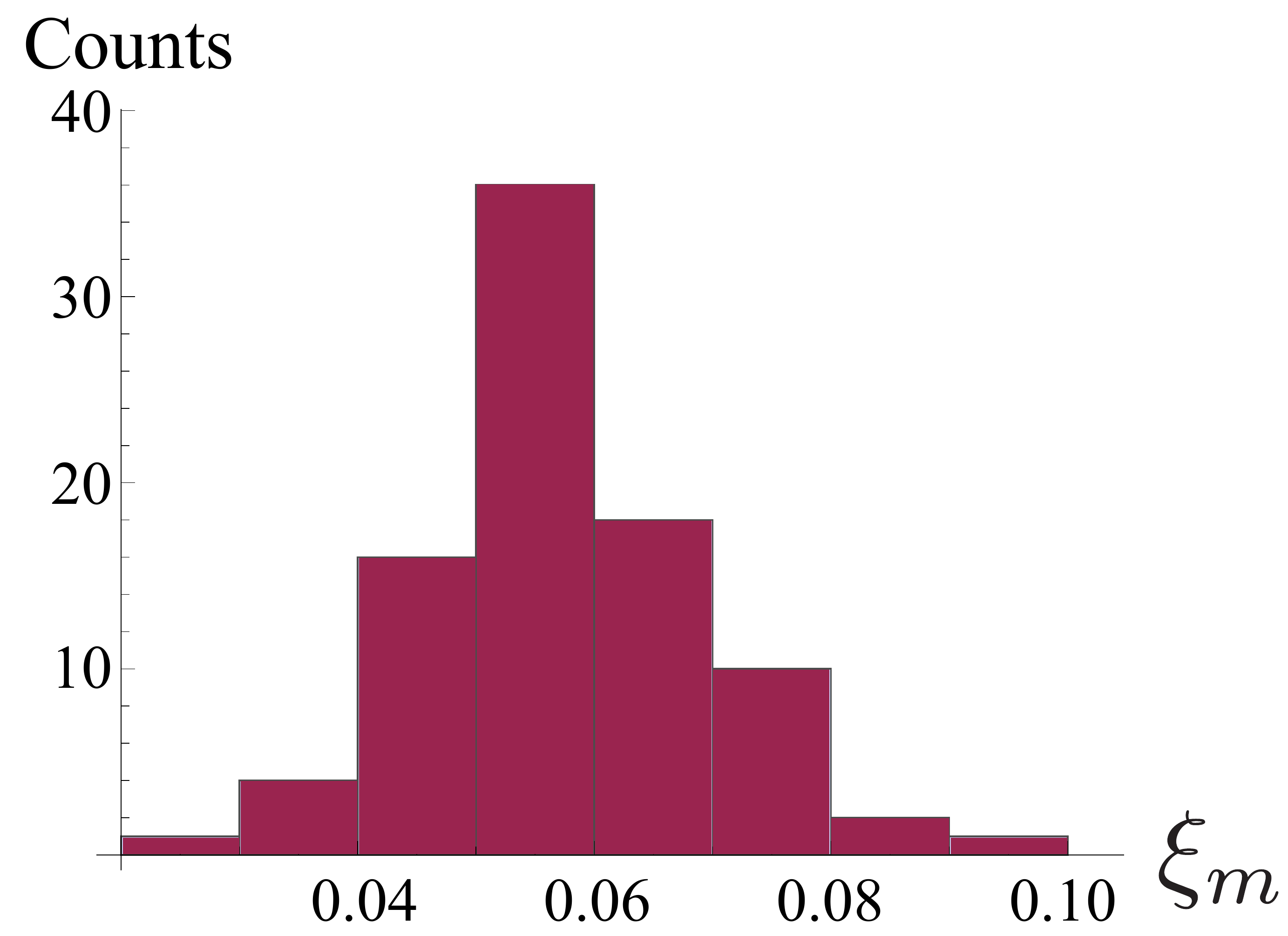}\includegraphics[width=0.5\columnwidth]{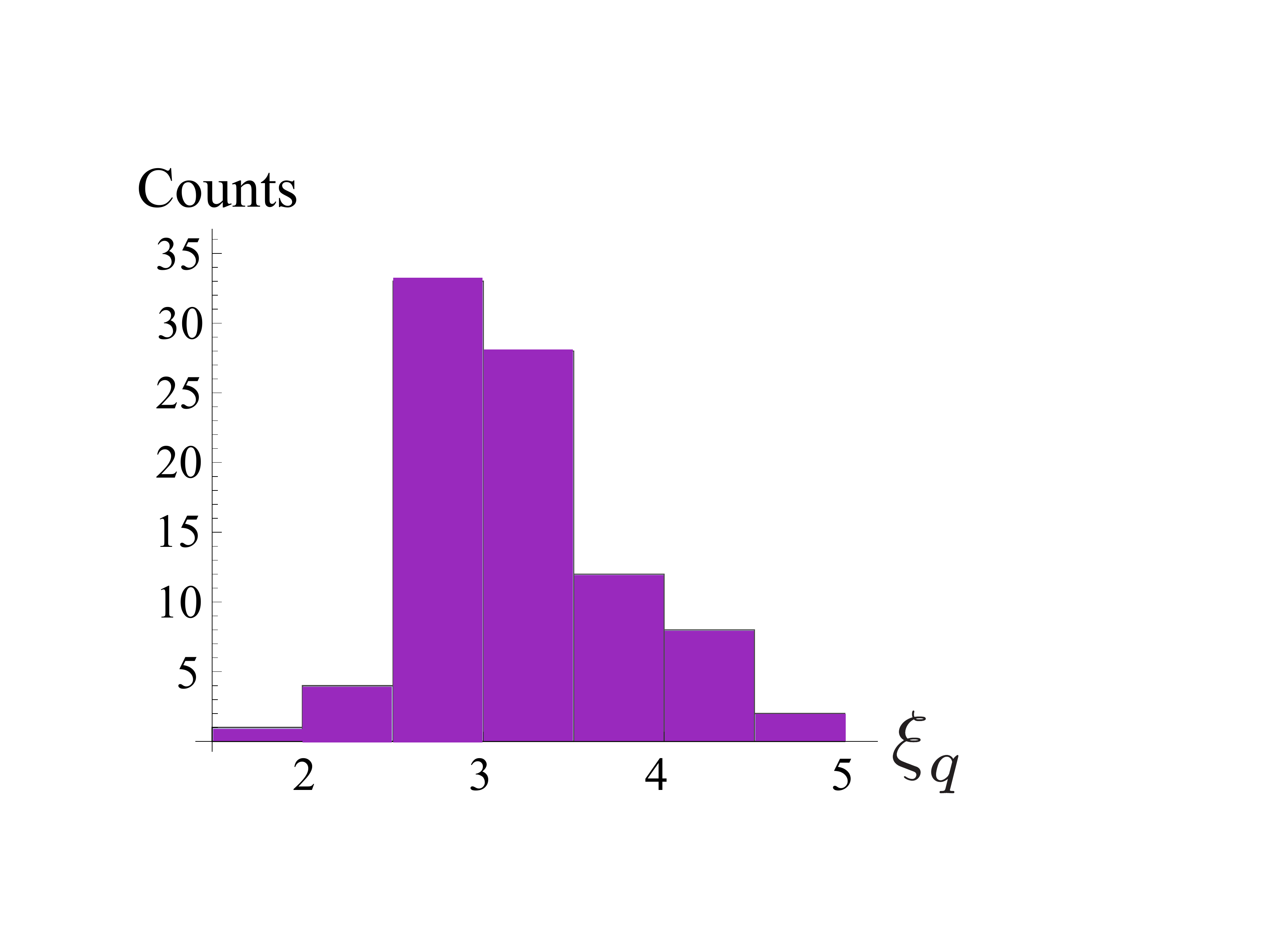} 
    \caption{(Color online) Distribution of values taken by the parameter $\xi_m$ [panel {\bf (a)}] and  $\xi_v$ [panel {\bf (b)}] obtained for 100 randomly-generated initial states of the BEC. Here we have taken the parameters used in panels {\bf (c)} and {\bf (d)} of Fig.~\ref{nm}. We have taken $t_0=0.07/\kappa$ and $t_1=0.8/\kappa$.}
  \label{ximxivNew}
\end{figure}

Finally, we discuss the generality of the method just described. We want this procedure to be valid for any initial state. To estimate the discrepancy between the measured value $\langle \hat n_1 \rangle_c$ and the real one, we can use the parameter
\begin{equation}
\xi_m=\frac{1}{t_1-t_0}\int_{t_0}^{t_1}|\langle \hat n_1 \rangle_c-\langle \hat n_1 \rangle|dt
\end{equation}
while, for the variance
\begin{equation}
\xi_q=\frac{1}{t_1-t_0}\int_{t_0}^{t_1}|\langle \hat n_1^2\rangle -\langle \hat n_1^2\rangle_c| dt
\end{equation}
where $t_0$ and $t_1$ are both times after the transient. In Figs.~\ref{ximxiv} and \ref{ximxivNew} we show histograms displaying the distribution of the values achieved by $\xi_{m,q}$  when 100 random initial states of the atomic system are prepared. Such atomic states are built as $\ket{r_i}=\sum^N_{n=0}c^i_n\ket{n,N-n}$ with $\{c^i_n\}$ a set of random complex numbers (such that $\sum^N_{n=0}|c^i_n|^2=1$) sampled uniformly for $i=1,..,100$. We have considered  both $R=\kappa$ (cf. Fig.~\ref{ximxiv}) and in the tunnelling-dominated regime (cf. Fig.~\ref{ximxivNew}). These values lie in a very narrow range with respect to the nominal values taken by ($\langle\hat n_1\rangle, \langle\hat n^2_1\rangle$) (cf. Fig.~\ref{nm}), thus showing the weak dependence of our results on the initial state of the system. We can thus claim the general validity of our approach, regardless of the dynamical conditions and the preparation of the system to probe.

\section{QET approach: determining the parameters of the Hamiltonian}
\label{QET}

The approach described in the previous Section relies on the one-to-one mapping of the information encoded in the atomic degrees of freedom into 
those of the probing cavity field. However, our strategy is not flexible enough to allow us to estimate other important parameter of the atomic Hamiltonian. In particular, the dynamics of the population imbalance between the wells is strongly dependent on the actual value of the parameters entering the Hamiltonian. It is thus interesting to determine precisely such values for a given situation. In what follows, we aim at providing an analysis of the precision with which the parameters of the Hamiltonian could be determined experimentally. Our study relies on the application of tools from (local) QET~\cite{QET} to the system of a driven cavity interacting with one of the wells of the atomic Josephson junction. 

In any estimation procedure the information about the quantity of interest is inferred from some suitable 
measurement performed on the system. Once the measurement has been 
chosen, an estimator is needed, i.e. a function from the data sample to the quantity  
of interest. Without specifying explicitly the parameter that we aim at estimating, we now go through a brief overview of the 
quantum parameter estimation theory to define the context and tools of our analysis.   

The variance $\mbox{Var}(\mu)$ of any unbiased estimator is 
lower-bounded, as stated by the Cram\'er-Rao inequality
\begin{equation}
\label{bound}
\mbox{Var}(\mu)\ge\frac{1}{MF(\mu)}
\end{equation}
with $M$ the number of measurements employed in the 
estimation and $F(\mu)$ the Fisher information relative to the parameter $\mu$. For measurements having a discrete set of outcomes, the Fisher information is 
defined as 
\begin{equation}
F(\mu)=\sum_{j}p_{j}(\partial_\mu\ln p_{j})^2=
\sum_{j}\frac{|\partial_\mu p_j|^2}{p_j},
\end{equation}
where $p_{j}$ represents the probability to get outcome $j$ from a
measurement performed over the probe state $\varrho(\mu)$.
On the other hand, for a continuous distribution of measurement outcomes, the above definition is changed by replacing the sum with an integral and $p_{j}\to p(x|\mu)$ with $p(x|\mu)$ 
the conditional distribution of obtaining the outcome $x$ at a set value of $\mu$. Quantum mechanically, such probabilities are calculated via the Born 
rule assuming the system at hand is in a state $\rho(\mu)$ determined by a set value of $\mu$. For the case of a discrete-measurement spectrum, which will be the one we will concentrate on in the rest of this manuscript, we thus have $p_j=\mathrm{Tr}[\rho(\mu)\hat\Pi_j]$, and the observable to 
be measured is generally  described by a positive operator valued 
measurement (POVM) built as $\{\hat\Pi_j : \hat\Pi_j\ge0, \sum_j\hat \Pi_j=\hat\openone\}$. Introducing the symmetric logarithmic derivative as the self-adjoint operator that satisfies the relation~\cite{parisReview}
\begin{equation}
\label{SLD}
\partial_\mu\rho(\mu)=\frac12[\hat L(\mu)\rho(\mu)+\rho(\mu) \hat L(\mu)],
\end{equation}
and optimising $F(\mu)$ over all possible quantum measurements leads us to the {\it quantum Fisher information} ${\cal H}(\mu)=\max_{\{\hat \Pi_j\}}F(\mu)$, which can be cast into the form  
\begin{equation}
{\cal H}(\mu)=\mathrm{Tr}[\rho(\mu) \hat{L}^2(\mu)].
\end{equation}
The \qfi is thus independent of the specific measurement strategy and leads to the extension of the Cram\'er-Rao bound to the quantum domain  
\begin{equation}
\mbox{Var}(\mu)\ge\frac{1}{M{\cal H}(\mu)},
\end{equation}
which embodies the ultimate limit to the precision of the estimate of $\mu$. Optimal quantum measurements correspond 
to POVMs whose $F(\mu)$ equals the \qfi\!.
\begin{figure*}[t]
\center{{\bf (a)}\hskip7cm{\bf (b)}}\\
\includegraphics[width=0.8\columnwidth]{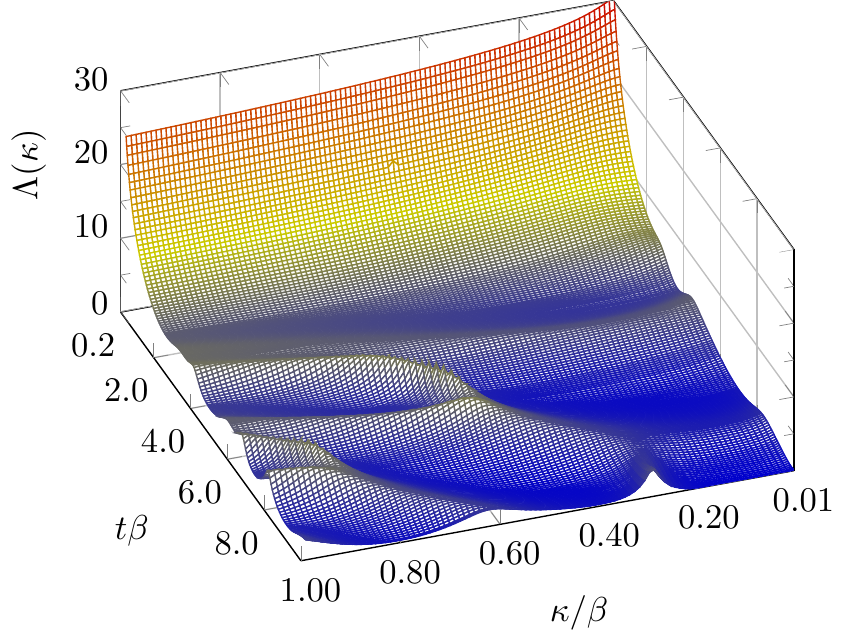}\includegraphics[width=0.8\columnwidth]{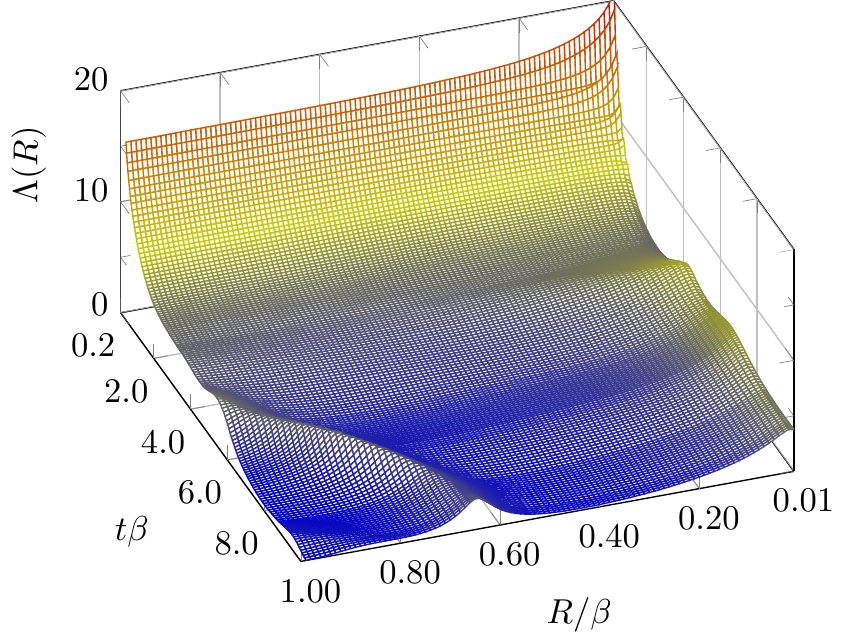}
\caption{(Color online) {\bf (a)} Single-parameter estimation of the self-interaction rate $\kappa$: we plot the single-parameter quantitity $\Lambda(\kappa)=\ln[{\cal H}^{-1}(\kappa)/\beta^2]$ at a set value of the tunnelling rate $R/\beta=0.5$, against the evolution time and the actual self-interaction energy $\kappa$. In panel {\bf (b)} we plot $\Lambda(R)=\ln[{\cal H}^{-1}(R)/\beta^2]$ against $R$ and $\beta t$ at $\kappa/\beta=0.5$. In both panels we have taken $\omega_a/\beta=0.1$, $\omega_c/\beta=0.1$, $\omega_p/\beta=0.1$, $\eta/\beta=0.1$ and $\gamma/\beta=1$.}
\label{singleparameter}
\end{figure*}
Eq.~\eqref{SLD} is a Lyapunov matrix equation whose general solution reads
\begin{equation}
\begin{aligned}
\hat L(\mu)&=2\int^\infty_0 dt\,e^{-\rho(\mu)t}\partial_\mu\rho(\mu)e^{\rho(\mu)t}\\
&=2\sum_{n,m}\frac{\langle\psi_m|\partial_\mu\rho(\mu)|\psi_n\rangle}{\rho_n(\mu)+\rho_m(\mu)}\ket{\psi_m}\bra{\psi_n},
\end{aligned}
\end{equation}
where we have used the spectral decomposition of the density matrix of the system $\rho(\mu)=\sum_{n}\rho_n(\mu)\ket{\psi_n}\bra{\psi_n}$. The
\qfi is correspondingly rewritten as~\cite{parisReview}
\begin{equation}
\label{explicit}
\begin{aligned}
{\cal H}(\mu)=\sum_p\frac{[\partial_\mu\rho_p(\mu)]^2}{\rho_p(\mu)}+2\sum_{m\neq n}\sigma_{mn}|\langle\psi_m|\partial_\mu\psi_n\rangle|^2
\end{aligned}
\end{equation}
with $\sigma_{nm}=2\left[\frac{\rho_n(\mu)-\rho_m(\mu)}{\rho_n(\mu)+\rho_m(\mu)}\right]^2$.
The first term in ${\cal H}(\mu)$ is the classical Fisher information of the distribution $\{\rho_n(\mu)\}$, while the second embodies the genuinely quantum part, which will be the focus of our attention
from this point on. In order to calculate such quantum contribution, we expand each eigenstate $\ket{\psi_n}$ over the orthonormal basis of Fock states $\{\ket{k}\}$ as
\begin{equation}
\ket{\psi_n}=\sum_k\psi_{nk}\ket{k},
\end{equation}
so that Eq.~\eqref{explicit} becomes ${\cal H}(\mu)={\cal H}_C(\mu)+{\cal H}_Q(\mu)$ with 
\begin{equation}
\begin{aligned}
{\cal H}_C(\mu)&=\sum_p\frac{[\partial_\mu\rho_p(\mu)]^2}{\rho_p(\mu)}~~~{\rm and}\\
{\cal H}_Q(\mu)&=\sum_{m\neq n}\frac{4\rho_n\left\|\sum_{kk'}\partial_\mu\left(\sum_l\rho_l(\mu)\psi_{lk}\psi^*_{lk'}\right)\psi^*_{mk}\psi_{nk'}\right\|^2}{(\rho_n+\rho_m)^2}.
\end{aligned}
\end{equation}

\begin{figure*}[ht]
\center{{\bf (a)}\hskip7cm{\bf (b)}}
\includegraphics[width=0.8\columnwidth]{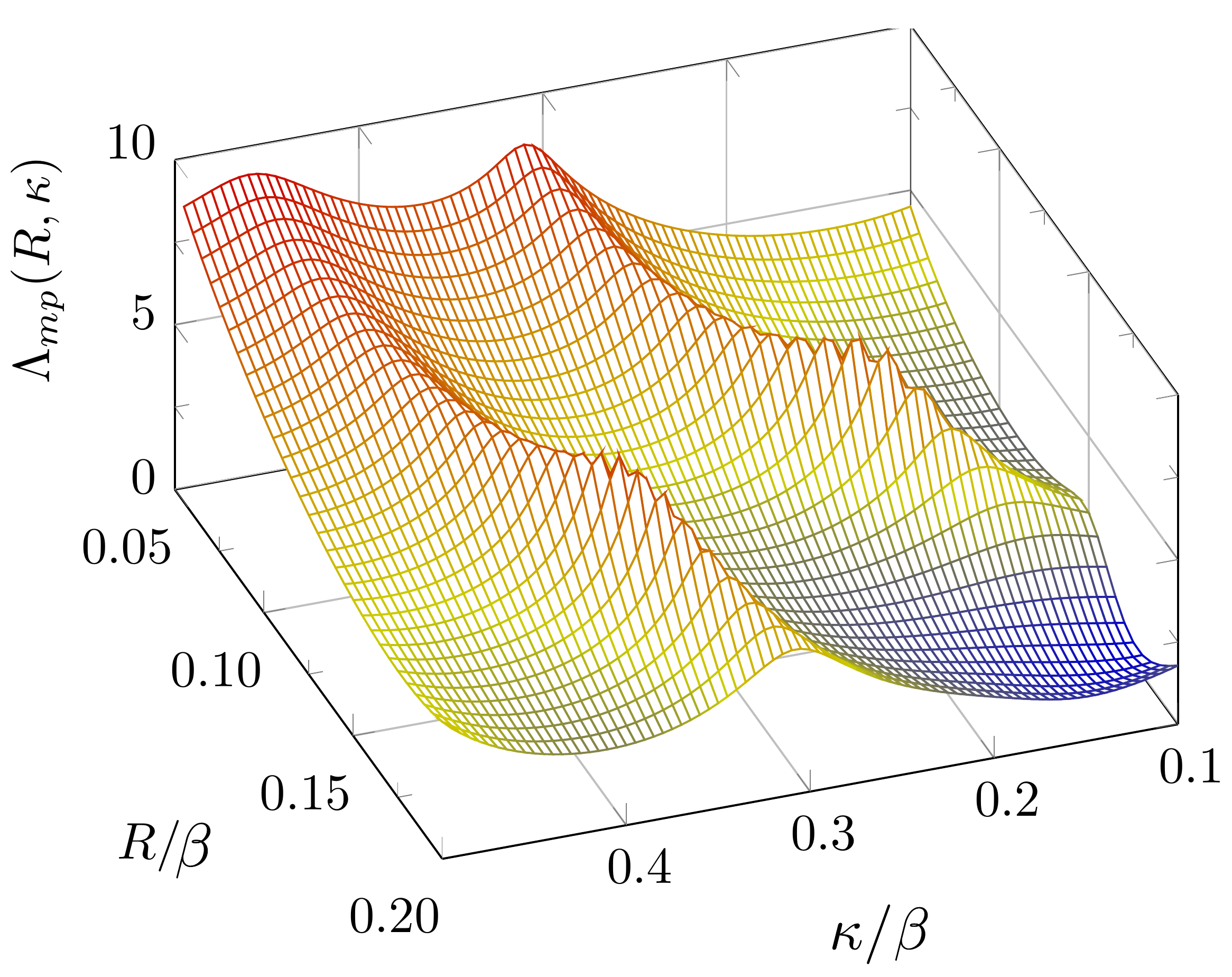}\includegraphics[width=0.8\columnwidth]{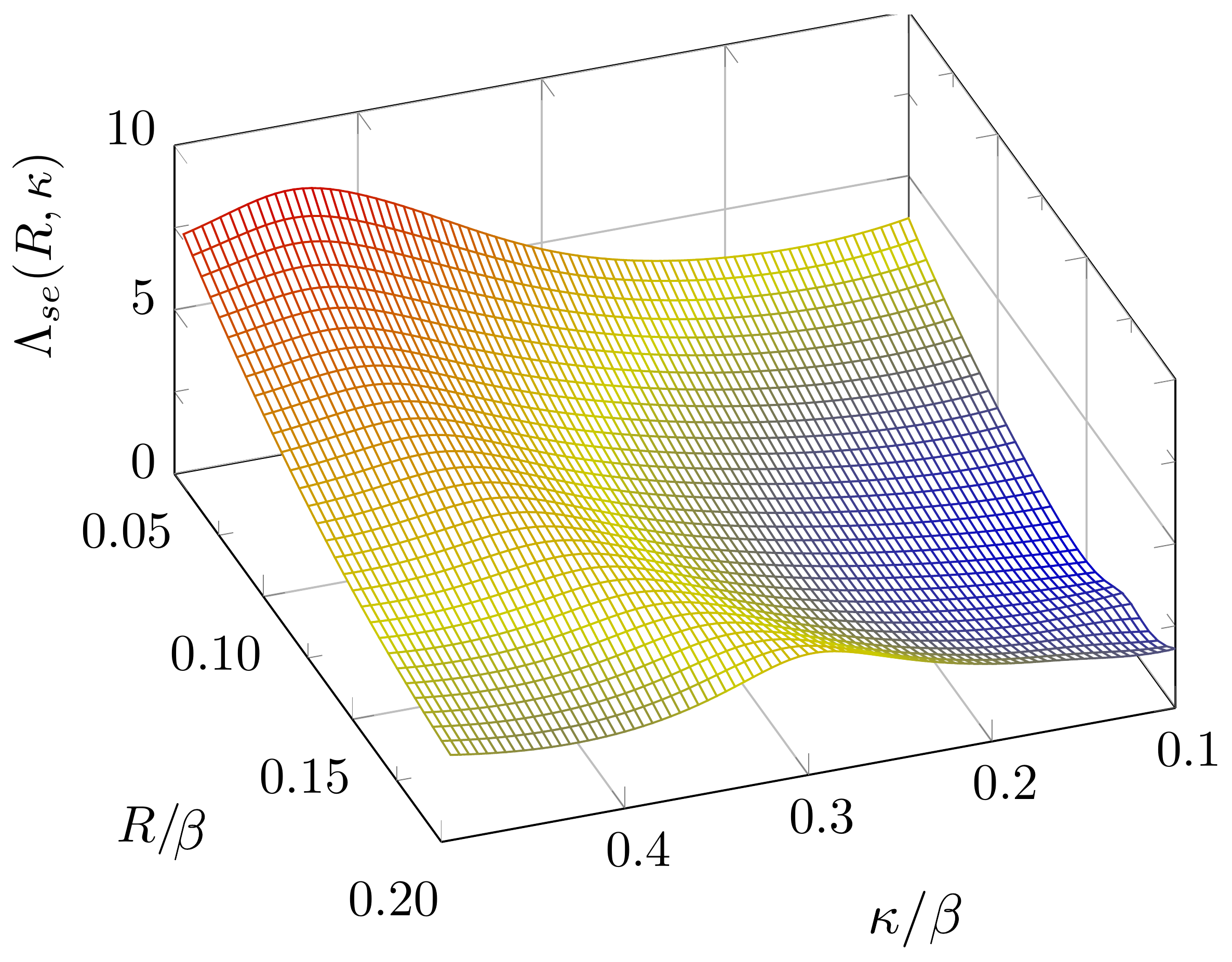}
\caption{(Color online) {\bf (a)} Multi-parameter estimation approach. {\bf (b)} Sequential estimation approach. In both panels we have used the parameters 
$\gamma/\beta=1$, 
$\eta/\beta=0.1$, 
$\omega_p/\beta=0.1$, 
$\omega_c/\beta=0.1$,
$\omega_a/\beta=0.1$. The evolution time has been set at $t=10/\beta$ and we have considered a two-atom initial state with mode $1$ initially fully populated and an empty cavity.}
\label{doubleparameter}
\end{figure*}

On the other hand, for a multi-parameter scenario, i.e. a situation where the state of the system under scrutiny depends on a set of parameters $\{\mu_1,\mu_2,..,\mu_d\}$, the formalism introduced above can be re-stated with the introduction of the parameter-specific symmetric logarithmic derivative
\begin{equation}
2\partial_{\mu_n}\rho(\mu_1,\mu_2,..,\mu_d)=\hat L_{\mu_n}\rho+\rho\hat L_{\mu_n}
\end{equation}
from which it is possible to define the \qfi matrix ${\bf H}$ with elements
\begin{equation}
{\cal H}_{np}=\text{Tr}\left[\rho(\mu_1,\mu_2,..,\mu_d)\frac{\hat L_{\mu_n}\hat L_{\mu_p}+\hat L_{\mu_p}\hat L_{\mu_n}}{2}\right].
\end{equation}
The quantum Cram\'er-Rao bound stated above is now replaced by the multi-parameter one that, in this paper, will be considered to be under the form
\begin{equation}
\sum_n\text{Var}(\mu_n)\ge\frac{1}{M}\text{Tr}[{\bm H}^{-1}].
\end{equation}
As before, $M$ is the number of measurements performed on the state of the system. While the single-parameter quantum Cram\'er-Rao bound is, in principle, always achievable through the design of an optimal measurement strategy, the multi-parameter counterpart is not, in general~\cite{vaneph}.

We now apply the frameworks discussed above to the problem of estimating the parameters of the Hamiltonian of the atomic system through indirect measurements performed on the field, which will be hereafter considered as a ``quantum probe". We will implement both a multi-parameter strategy and a ``sequential" estimation approach, where $M$ measurements are used to estimate the tunnelling rate while a separate set of $M$ additional measurements are instrumental to the estimation of the self-interaction strength. 

\begin{figure*}[ht]
\center{{\bf (a)}\hskip6.5cm{\bf (b)}}
\includegraphics[width=0.8\columnwidth]{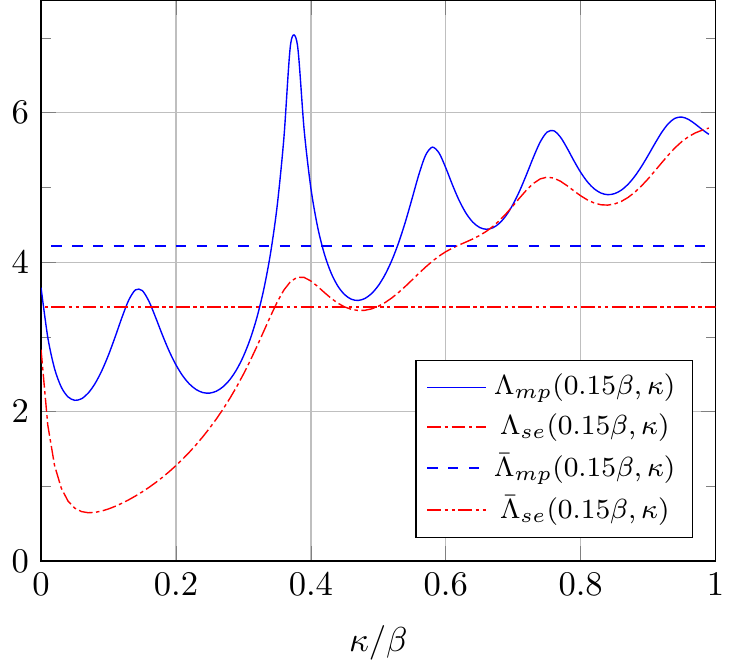}\includegraphics[width=0.8\columnwidth]{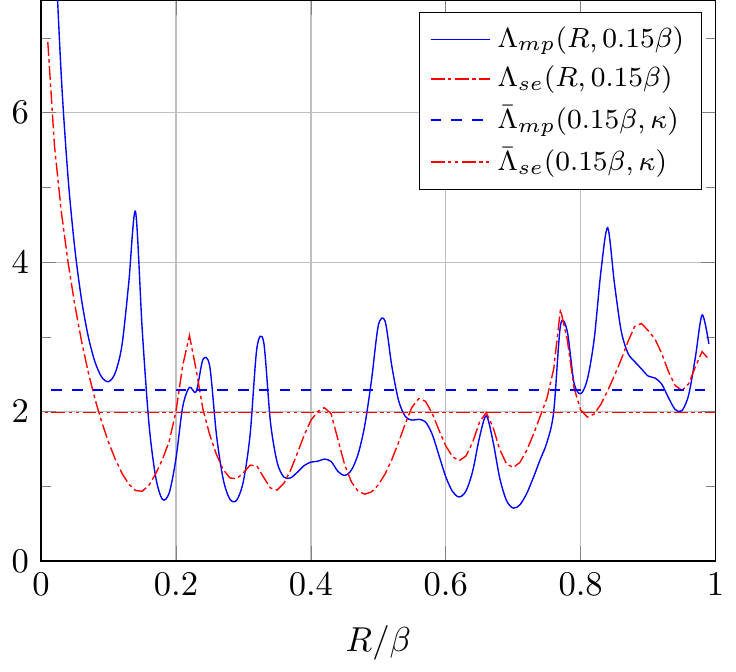}
\caption{(Color online) Multi-parameter vs. sequential estimation approach. We compare $\Lambda_{mp}(R,\kappa)$ and $\Lambda_{se}(R,\kappa)$ in various parameter regimes. In panel {\bf (a)} we plot the two figures of merit at $R/\beta=0.15$ and investigate the estimation performance through the two strategies as the value of $\kappa$ is changed. Panel {\bf (b)} displays a similar comparison performed against the tunnelling rate $R$ at $\kappa/\beta=0.15$. The straight lines ($\bar{\Lambda}_{se}$, $\bar{\Lambda}_{mp}$) show the average of the two figures of merit over the range of values taken by $\kappa$ [panel {\bf (a)}] and $R$ [panel {\bf (b)}]. In both panels we have taken $\gamma/\beta=1$, 
$\eta/\beta=0.1$, 
$\omega_p/\beta=0.1$, 
$\omega_c/\beta=0.1$,
$\omega_a/\beta=0.1$. The evolution time has been set at $t=10/\beta$ and we have considered a two-atom initial state with mode $1$ initially fully populated and an empty cavity.}
\label{comparison}
\end{figure*}

In order to proceed along such lines, we need to determine the state of the field after its interaction with the atomic system in the double-well. Although the accessible part of the electromagnetic signal is what leaks out of the probing cavity, considering the out-coming field simply adds shot noise to the estimation procedure, as it is straightforward to see by considering standard cavity input-output relations. We thus concentrate on the dynamics of the intra-cavity field, without any loss of generality, and study the maximum precision with which we can estimate the tunnelling rate $R$ or the self-interaction rate $\kappa$. The dynamical model that we aim at solving is thus 
\begin{equation}
\begin{aligned}
\partial_t\rhor=&-i[\hat H_A+\hat H_I,\rhor]-i\Delta_C[\hat a^\dag\hat a,\rhor]\\
&-i\eta[\hat a^\dag+\hat a,\rhor]-\gamma D[\hat a,\rhor]
\end{aligned}
\end{equation}
with $\rhor$ the density matrix of the system in a frame rotating at the frequency of the pump. This equation is tackled by projecting it onto the elements of a number-state basis of elements $\ket{n1,N-n1,n_a}$. Here $\ket{n_1}$ [$\ket{N-n_1}$] is a state with $n_1$ [$N-n_1$] atoms in the well coupled [not coupled] to the cavity, and $\ket{n_a}$ is a similar state for the field. This allows us to get nested Bloch-like equations that have then been solved numerically. By considering the decomposition
\begin{equation}
\rhor=\sum {\cal C}_{n_1,n_a}^{m_1,m_a}\ket{n_1,N-n_1,n_a}\bra{m_1,N-m_1,m_a}
\end{equation}
with the sum being extended over all values of the indices and $ {\cal C}_{n_1,n_a}^{m_1,m_a}=\bra{m_1,N-m_1,m_a}\rhor\ket{n_1,N-n_1,n_a}$, we get the reduced field state (in the Schr\"odinger picture)
\begin{equation}
\rho_C=\sum_{n_1,n_a,m_a}e^{-i(n_a-m_a)\omega_p t}{\cal C}_{n_1,n_a}^{n_1,m_a}\ket{n_a}\bra{m_a},
\end{equation}
which is then used to apply the quantum parameter estimation framework introduced above.  

Before attacking the problem of estimating both the key parameters of the system's Hamiltonian, namely $R$ and $\kappa$, we construct a useful benchmark by by addressing a single-parameter estimation problem where we fix one of such parameters and evaluate the accuracy of estimation of the other one. This study is reported in Fig.~\ref{singleparameter}. Panel {\bf (a)} shows the behavior of the quantity $\Lambda(\kappa)=\ln[{\cal H}^{-1}(\kappa)/\beta^2]$ at $R=0.15\beta$, which gives information on the value of the minimum variance associated with the estimation of the self-interaction energy. While it should be clear that the value of $\Lambda(\kappa)$ depends on the instant of time at which the estimation procedure is implemented, we highlight the fact that different estimation performances might be achieved as the actual value of $\kappa$ is changed. We thus show such parameter against the dimensionless evolution time $\beta t$ and the value taken by $\kappa$ (in units of $\beta$). A similar analysis for the case of the estimation of the tunnelling rate $R$ is reported in Fig.~\ref{singleparameter} {\bf (b)}, where we study $\Lambda(R)=\ln[{\cal H}^{-1}(R)/\beta^2]$ in the $(\beta t,R/\beta)$ space at $\kappa/\beta=0.15$. Only a very weak dependence on the actual value taken by the parameter to be estimated is displayed by $\Lambda(\mu)~~(\mu=R,\kappa)$, which is instead very much dependent on the actual time of the evolution. Short times, associated with the transient part of the evolution of the system, correspond to very large values of the variance and thus a poor estimation. On the other hand, by approaching the time-asymptotic regime, where the system reaches a quasi steady-state, an accurate estimation of any of the two parameters is possible, in line with the expectations gathered from the analysis based on our light-matter mapping. 

We now pass to address a somehow different question related to the possibility to estimate {\it both} the tunnelling and self-interaction rates, and thus characterize completely the non-trivial part of the Hamiltonian of the system. As anticipated above, when dealing with such a multi-parameter problem, one can adopt either a parallel or a sequential approach. The performance of the two approaches, in terms of uncertainty associated with the estimation process, might differ substantially. In order to gather a quantitative comparison, in Fig.~\ref{doubleparameter} we plot  the natural logarithm of the minimum variance $\text{Var}(\mu)$ achieved in the process of estimating $\kappa$ and $R$ for, respectively, the multi-parameter strategy and the sequential strategy. Refs.~\cite{Dariano} have discussed efficient ways for the estimation of the parameters of a generic quadratic Hamiltonian for two bosonic modes. For the multi-parameter (sequential) strategy, such variance is determined by the quantity $\Lambda_{mp}(R,\kappa)=\ln[({\cal H}^{-1}(R)+{\cal H}^{-1}(\kappa))/(2\beta^2)]$ ($\Lambda_{sq}(R,\kappa)=\ln[({\cal H}^{-1}(R)+{\cal H}^{-1}(\kappa))/\beta^2]$). The factor 2 introduced in $\Lambda_{mp}(R,\kappa)$ is necessary to compare fairly with $\Lambda_{se}(R,\kappa)$. 

The dynamical nature of the problem that we are addressing, then makes the study quite rich and complex. In light of what we have found through the analysis of the single-parameter estimation process, our study (not reported here) shows that both $\Lambda_{mp}(R,\kappa)$ and $\Lambda_{se}(R,\kappa)$ change dynamically,  providing smaller uncertainties at larger evolution times. Therefore, in order to minimise the effects of the transient dynamics, we decide to focus on the long-time limit and take  $\beta t=10$ in all of the quantitative studies presented here. Fig.~\ref{doubleparameter} is well representative of the results that we have gathered, encompassing both the dynamical regimes that we have addressed throughout our study, i.e. the case of $R\sim\kappa$ and the tunnelling dominated configuration. The inspection of Fig.~\ref{doubleparameter} reveals that both the multi-parameter and  sequential approaches give rise to a rather rich behavior of the minimum variance associated with the parameters being estimated. The parameters entering our simulations should be chosen properly, as the sensitivity of the estimation indeed strongly depends on our working point. 

In general, the best option between the sequential and multi-parameter approach depends strongly on the values of $\kappa$ and $R$ that we aim at estimating. While Fig.~\ref{comparison}  {\bf (a)} illustrates a case where a multi-parameter estimation strategy is almost always inferior to the sequential approach, panel {\bf (b)} explores a working point where this is not always the case, thus making any general prediction very difficult. However, we can abandon such a {`point-by-point'} analysis in favour of an {`integrated'} evaluation of the estimation performance. Our line of thoughts would be the following: in a given estimation problem, it is unlikely to be completely ignorant of the order of magnitude taken by the parameter that we aim at determining. Differently, it is reasonable to expect that pre-available information (for instance on the actual working conditions under which an experiment would be run) could be used to gauge the plausible range of values that it could take. The question that we aim at addressing, in this case, would be: {\it For unknown parameters $\{\mu_j\}$ lying in the regions $\{\Omega_{\mu_j}\}$, which estimation strategy (either multi-parameter or sequential) is more advantageous, on average?} In our case, a quantitative assessment of such a problem could come from the consideration of the average value taken by $\Lambda_{mp}(R,\kappa)$ and $\Lambda_{se}(R,\kappa)$. This is what is shown by the dashed lines in Fig.~\ref{comparison}. Clearly, the consideration of such an average figure of merit, although making us lose the details of the point-to-point behavior of the minimum variances associated with the estimates of $R$ and $\kappa$, provides useful information: a sequential approach turns out to be more advantageous than the multi-parameter strategy, which delivers a consistently larger value of the associated minimum variance. Such a behavior is not restricted to the working point used in Fig.~\ref{comparison} but turns out to be consistent across the range of values considered in Fig.~\ref{doubleparameter}.

\section{Conclusions and outlook}
\label{conc}

We have proposed a QET-based approach to the determination of key parameters in the dynamics of an atomic system loaded into a two-well potential. Our technique makes use of a {\it local} quantum probe embodied by the field of an optical cavity that is coupled only to one of the wells of the potential. We have shown that a variety of methods can be applied in order to estimate crucial features of the dynamics of the double well, from the population imbalance between the wells to the actual on-site self interaction energy and tunnelling rate characterizing the Hamiltonian of the atoms loaded into the potential. By evaluating the quantum Fisher information associated with the specific problem at hand, we have been able to determine the quantum-limited precision with which is possible to estimate the parameters of the problem's Hamiltonian in both a sequential and a multi-parameter estimation approach. While the best strategy to follow in order to achieve such ideal estimates appears to depend crucially on the actual dynamical working point at hand, our work opens up a series of routes that will be explored in our forthcoming work, from the estimate of the temperature of the atoms loaded in the wells to the explicit quantification of the rate at which the atomic system equilibrates. 

\acknowledgments 

MZ and JPS thanks the Centre for Theoretical Atomic, Molecular, and Optical Physics, Queen's University Belfast for hospitality during the early stages of this work. MZ acknowledges financial support from Universit\'a degli Studi di Palermo under the PERFEST 2011 initiative and from the National Research Foundation and Ministry of Education in Singapore. The authors acknowledge financial support from the UK EPSRC (EP/L005026/1, EP/G004579/1), the John Templeton Foundation (grant ID 43467), the Italian MIUR-PRIN 2010/2011, the EU Collaborative Project TherMiQ (Grant Agreement 618074). JPS acknowledges FAPESP (GrantNo. 2011/09258-5). JPS, FLS and MP are supported by the CNPq ``Ci\^{e}ncia sem Fronteiras'' programme through the ``Pesquisador Visitante Especial'' initiative (grant nr. 401265/2012-9). FLS is a member of the Brazilian National Institute of Science and Technology of Quantum Information (INCT-IQ) and acknowledges partial support from CNPq (grant nr. 308948/2011-4).

\end{document}